\documentclass[letterpaper,english,reprint, aps, longbibliography]{revtex4-1}
\usepackage[T1]{fontenc}
\usepackage[latin9]{inputenc}
\usepackage{amsmath}
\usepackage{amssymb}

\makeatletter

\pdfpageheight\paperheight
\pdfpagewidth\paperwidth

\providecommand{\tabularnewline}{\\}

\makeatother

\usepackage{babel}
\begin{document}
\title{Semigroup approach to the sign problem in quantum Monte Carlo simulations}
\author{Zhong-Chao Wei}
\email{zcwei@thp.uni-koeln.de}

\affiliation{Institute for Theoretical Physics, Department of Physics, University
of Cologne, Cologne 50937, Germany}
\date{\today}
\begin{abstract}
We propose a framework based on the concept of the semigroup to understand
the fermion sign problem. By using properties of contraction semigroups,
we obtain sufficient conditions for quantum lattice fermion models
to be sign-problem-free. Many previous results can be considered as
special cases of our new results. As a direct application of our new
results, we construct a class of sign-problem-free fermion lattice
models, which cannot be understood by previous frameworks. This framework
also provides an interesting aspect in understanding related quantum
many-body systems. We establish a series of inequalities for all the
sign-problem-free fermion lattice models that satisfy our sufficient
conditions.
\end{abstract}
\pacs{02.70.Ss, 71.10.Fd, 71.27.+a}
\keywords{sign problem, semigroup, quantum Monte Carlo}
\maketitle

\section{Introduction}

Understanding interacting many-body systems remains a great challenge
in current physics research. The quantum Monte Carlo (QMC) method
is an important numerical method for this purpose\citep{ceperley_path_1995,foulkes_quantum_2001,gull_continuous-time_2011,carlson_quantum_2015}.
It contains a class of stochastic algorithms based on sampling over
different configurations according to some sampling weights derived
from the model. However, for many quantum models it is often extremely
difficult to express the quantum partition function or expectation
values of physical variables in terms of efficiently computable, non-negative
real sampling weights. This obstacle, which often hampers the efficiency
of QMC simulations seriously, is called the sign problem. It prevents
us from effectively getting numerical results for large systems at
low temperature.

Specifically speaking, for auxiliary field quantum Monte Carlo (AFQMC)
type algorithms\citep{blankenbecler_monte_1981,hirsch_two-dimensional_1985}
that are frequently used in condensed-matter physics, nuclear physics,
and cold atoms, for each configuration of auxiliary fields the contribution
to the partition function can be expressed by the determinant resulting
from the fermionic Gaussian integral, which can be computed efficiently.
Unfortunately, in general, a fermionic Gaussian integral is not necessarily
a real number, even less a non-negative real number. For fermion lattice
models, the sign problem will lead to an exponential growth of total
computational cost as the volume of the system and the inversed temperature
get larger\citep{loh_sign_1990,chandrasekharan_meron-cluster_1999},
if one wants to retain the same numerical accuracy.

Despite the fact that a general unbiased solution to the sign problem
is either non-existent or elusive by its very nature\citep{troyer_computational_2005},
a lot of physically interesting models have been shown to be sign-problem-free,
which is of great significance to practical numerical studies. For
AFQMC and some related methods, a few general frameworks have been
proposed to understand sign-problem-free interacting fermion systems.
There have been approaches based on the Kramers time-reversal invariance\citep{lang_monte_1993,wu_exact_2003,wu_sufficient_2005},
the fermion bag\citep{chandrasekharan_fermion_2010,huffman_solution_2014},
the Majorana quantum Monte Carlo\citep{li_solving_2015}, the split
orthogonal group\citep{wang_split_2015}, the Majorana reflection
positivity\citep{wei_majorana_2016}, and the Majorana time-reversal
symmetries\citep{li_majorana-time-reversal_2016}. Each approach has
unified a class of sign-problem-free fermion models and brought additional
examples of sign-problem-free QMC simulations.

In this work, we propose an alternative approach to construct fermion
models without the sign problem. We observe that semigroup structures
arise naturally from imaginary-time evolutions; this observation is
made explicit after introducing auxiliary fields. The semigroup is
generated by multiplication of exponentials of fermionic quadratic
operators. It is not necessarily a group, for the inverse elements
of those exponentials may not appear in the calculation. An important
particular case is when each element of the semigroup has non-negative
trace, then the QMC sampling weights are exactly the traces. This
fact serves as the stating point of our approach.

A semigroup is a set with element multiplication that satisfies the
associative law. Compared with the concept of a group, an inverse
does not necessarily exist for each element in a semigroup. Semigroups
appear frequently in different areas of theoretical research. Every
group is also a semigroup. In quantum mechanics, the quantum dynamical
semigroup\citep{alicki_quantum_2007} is employed to study the time
evolution of open quantum systems, where the concept of the semigroup
reflects the irreversiblility of time for the concerned physical processes.
In quantum field theory and statistical physics, the renormalization
group(RG)\citep{wilson_renormalization_1983} is actually more like
a semigroup than a group, due to the loss of information during RG
transformations.

In this work, we are mainly concerned with a special kind of Lie semigroup
called the contraction semigroup. We construct two kinds of contraction
semigroups. When the parameter region is contained in such semigroups,
the fermionic Gaussian integral is always real non-negative. As a
result, the related AFQMC calculations do not have any sign problem.

The currently existing approaches mentioned above appear different
and unrelated at first glance, but now they can be unified in this
framework. The Kramers time-reversal invariance leads to Kramers pairs
of eigenvalues of matrices, which results in the non-negativity of
fermion determiants\citep{wu_sufficient_2005}. In Ref.~\citep{wang_split_2015},
the relation between the split orthogonal group and sign-problem-free
models has been revealed using some inequality for group elements.
Those results have been extended by recent studies\citep{wei_majorana_2016,li_majorana-time-reversal_2016}.
We show that all those approaches based on consideration of symmetries
are related to subgroups of the semigroups considered here. We also
explain that in the context of the AFQMC sign problem, the condition
of Majorana reflection positivity\citep{wei_majorana_2016} is actually
equivalent to one of the two kinds of contraction semigroups treated
in this work. In short, to our best knowledge, all known sign-problem-free
fermion lattice models used in simulation approaches based on semigroups
can be understood in our framework.

Our results open up new possibilities to sign-problem-free Monte Carlo
simulations. We construct a kind of interacting fermion lattice model
which involves the pairing term, the Kramers time-reversal invariant
hopping term, and the interaction term. This class of model is sign-problem-free,
which could not be explained in previous frameworks.

We believe that this framework will find more applications in both
numerical and analytical studies\citep{grossman_robust_2023}. To
illustrate the latter case, we establish certain inequalities for
the expectation values of physical observables in many-body systems.

\section{Problem Setting}

In AFQMC algorithms for interacting fermion lattice models, interaction
terms are decoupled by auxiliary fields into fermionic quadratic forms\citep{blankenbecler_monte_1981,hirsch_discrete_1983,hirsch_two-dimensional_1985}.
After integrating out the fermion degrees of freedom, one will obtain
an action in terms of auxiliary fields. One can treat this action
with random sampling numerically. The sampling weight for each configuration
of auxiliary fields usually has the form\citep{lang_monte_1993,klich_note_2014}
\begin{eqnarray}
p= &  & \ \mathrm{tr}\left(e^{h_{1}}e^{h_{2}}\cdots e^{h_{k}}\right)\nonumber \\
= &  & \ \det\left(I+e^{A_{1}}e^{A_{2}}\cdots e^{A_{k}}\right)^{1/2}.\label{eq:Trace}
\end{eqnarray}

The expression of sampling weight in Eq.~(\ref{eq:Trace}) is the
main object of study in this work. Here $h_{i}=\gamma^{T}A_{i}\gamma/4$
($i=1,\dots,k$) denotes a set of fermionic quadratic forms. They
come from both single-body terms and auxiliary field decoupling of
interaction terms, and depend on the configuration of auxiliary fields.
$\gamma_{n}$ ($n=1,\dots,2N$) are Majorana fermion operators, which
satisfy the anticommutation relations $\left\{ \gamma_{l},\ \gamma_{m}\right\} =2\delta_{lm}$.
The Majorana fermion basis is used for convenience; it is equivalent
to the complex fermion basis with $N$ species. Unless we specify
a particular example, in principle $h_{i}$ could contain an arbitrary
particle number conserving part and an arbitrary pairing part. That
is the equivalent of saying that the coefficients $A_{i}=-A_{i}^{T}$
could be arbitrary skew-symmetric complex matrices.

If we do not put any restrictions on $h_{i}$, the fermionic Gaussian
integral $p$ could be non-positive, due to both the complex nature
of the coefficients and the two-valuedness of the spin representation.
Under those circumstances, statistical sampling methods may fail to
obtain desired physical quantities with useful accuracy at reasonable
cost. This is the so-called sign problem in AFQMC methods.

In practical calculations, the possible forms of $e^{h_{i}}$ are
given by the quantum partition function. They could come from both
the single-body term in the Trotter-Suzuki decomposition and the Hubbard-Stratonovich
(HS) transformations for interaction terms. Their inverse elements
$e^{-h_{i}}$, however, are not necessarily involved in any fermionic
Gaussian integrals\footnote{The single-body term in the Hamiltonian could be a typical example.}.
By taking products along the imaginary-time axis, they form a semigroup,
with elements representing different sorts of paths of the partition
function. This observation allows us to study the sign problem in
terms of semigroups.

Furthermore, if a Lie semigroup $S\subset Spin\left(2N,\mathbb{C}\right)$
has the property $p\geq0$ for all its elements, then it corresponds
to a class of sign-problem-free fermion models. In the following sections,
we show that two specific kinds of Lie semigroups indeed possess this
good property.

\section{Definitions And Useful Facts}

We list some basic definitions and facts before going into details.
For general mathematical accounts of Lie semigroups, the reader may
refer to Refs.\ \citep{hilgert_lie_1993,hofmann_semigroups_2011}.

For any square complex matrix $X$, consider an antilinear symmetry
operation $X\mapsto\eta X^{\dagger}\eta$ given by a Hermitian matrix
$\eta=\eta^{\dagger}$ with $\eta^{2}=I$, together with Hermitian
conjugation. We say that the matrix is $\eta$-Hermitian or $\eta$-anti-Hermitian,
respectively, if it is invariant or changes sign under this transformation.
All the square complex matrices $X$ with property $\eta X+X^{\dagger}\eta\leq0$
generate a Lie semigroup by taking exponentials and element products.
Semigroups of this kind are called contraction semigroups. Equivalently,
one can say that the contraction semigroup consists of all the square
matrices $g$ that satisfy $g^{\dagger}\eta g-\eta\leq0$. They contract
the ``length'' of any vector given by the metric $\eta$. Similarly,
one can define the expansion semigroup by inverting the direction
of the inequality. We will work on the contraction semigroup and leave
the expansion case to the reader.

Obviously, the contraction semigroup defined above has the $\eta$-unitary
group as its maximal subgroup, which is generated by $\eta$-anti-Hermitian
matrices. Each element $g$ in the contraction semigroup possesses
a polar decomposition $g=g_{U}\exp\left(X_{0}\right)$, where $X_{0}$
is $\eta$-Hermitian and $\eta X_{0}\leq0$, and $g_{U}$ belongs
to the $\eta$-unitary group, i.e., $g_{U}^{\dagger}\eta g_{U}=\eta$.
The set of $X_{0}$ forms an invariant cone under adjoint action of
the $\eta$-unitary group.

Specially, let us consider strict contraction elements, which remain
strict contractions when multiplied by any semigroup elements. In
the strict contraction case $g^{\dagger}\eta g-\eta<0$, which implies
that $\eta X_{0}<0$ and $g$ cannot have eigenvalues of magnitude
$1$. This means $\det\left(I+g\right)\neq0$, which we use in the
following section to construct sign-problem-free semigroups.

\section{Sign-problem-free Semigroups}

Let us give the outline of the discussions in this section. First,
we restrict the range of parameters by an antilinear symmetry to make
the sampling weight $p$ real. Then we observe that $p$ never vanishes
for strict contraction elements inside some contraction semigroups,
while the nonstrict contraction elements can be viewed as some limit
of strict contraction elements. These two conditions together ensure
that $p$ is non-negative as a continuous function of the coefficients.

Each condition requires a definition of antilinear involution for
complex skew-symmetric matrices. Consider any complex skew-symmetric
matrix $A$. Adopting the Majorana fermion basis, it is natural to
assume that those two operations are expressed by real orthogonal
transformations acting on $A$, $J_{1}$, and $J_{2}$, respectively,
along with the complex conjugation.

First, we assume the complex skew-symmetric matrices are fixed under
the operation $A\mapsto J_{1}^{T}\bar{A}J_{1}$. Here $J_{1}$ could
be either symmetric $J_{1}^{2}=I_{2N}$ or skew-symmetric $J_{1}^{2}=-I_{2N}$.
It is easy to see that $p$ is real under this assumption.

Second, to define a contraction semigroup, $J_{2}$ should be chosen
to be either symmetric or skew-symmetric, so that $J_{2}$ or $iJ_{2}$
can serve as the aforementioned indefinite metric $\eta$. The coefficient
matrices that are not changed under the transformation $A\mapsto J_{2}^{T}\bar{A}J_{2}$
generate the maximal subgroup of the contraction semigroup. Meanwhile,
elements in the invariant cone change sign under this operation, $iJ_{2}A=-i\bar{A}J_{2}\leq0.$
Since $A$ is skew-symmetric, if $J_{2}$ were symmetric, the invariant
cone would be trivial, i.e., it contains only zero element. Therefore
we have to assume $J_{2}$ to be skew-symmetric. According to Eq.\ (\ref{eq:Trace}),
$p$ is nonzero for any such defined strict contraction element, because
the matrix inside the determinant does not have zero eigenvalues.

Finally, we have to check the consistency of the two conditions. The
resulting invariant cone should satisfy both constraints given by
$J_{1}$ and $J_{2}$. However, in order to make our argument stand,
we have to ensure that the resulting invariant cone always contains
strict contraction elements. This cannot be achieved by an arbitrary
choice of $J_{1}$ and $J_{2}$\footnote{For example, the cone is trivial when $\left[J_{1},\ J_{2}\right]=0$. }.
Under the current assumption, the only possibility is that $J_{1}$
and $J_{2}$ satisfy the anticommutation relation $\left\{ J_{1},\ J_{2}\right\} =0$.
See the Supplemental Material\footnote{See Supplemental Material at {[}URL will be inserted by publisher{]}
for three lengthy proofs which are not included in the main text.} for more detailed arguments. 

Now we have two sign-problem-free semigroups on which $p\geq0$, and
they are defined by
\begin{eqnarray}
J_{1}^{T}AJ_{1} & = & \bar{A},\label{eq:RealStructure}\\
i\left(J_{2}A-\bar{A}J_{2}\right) & \leq & 0.\label{eq:ContractionCondition}
\end{eqnarray}
$J_{1}$ and $J_{2}$ are two anti-commuting, real orthogonal matrices.
While $J_{1}$ could be symmetric or skew-symmetric, $J_{2}$ should
be skew-symmetric. If all $A_{i}$ matrices in Eq.~(\ref{eq:Trace})
satisfy the conditions above, the corresponding quantum Monte Carlo
simulations will be sign-problem-free.

Throughout the above discussions we do not require the Hermitian condition
of Majorana fermion operators $\gamma_{n}=\gamma_{n}^{\dagger}$.
Instead, the anticommutation relations for Majorana fermion operators
are preserved under complex orthogonal transformations of Majorana
fermion operators. Therefore, the condition for positive trace given
above also holds for this complex orthogonal generalization of the
Majorana fermion basis.

\section{Applications}

Equations~(\ref{eq:RealStructure}) and ~(\ref{eq:ContractionCondition})
constitute the main result of this work. They cover all the results
on sign-problem-free QMC simulations of fermion lattice models known
to us. They are classified and listed in Table~\ref{tab:Classification}
and are discussed type by type in this section.
\begin{table}

\caption{\label{tab:Classification}Classification of sign-problem-free quantum
lattice fermion models.}
\begin{tabular}{|c|c|c|c|}
\hline 
Cases & $J_{1,a}^{2}=I$, $J_{1,b}^{2}=-I$ & $J_{1}^{2}=I$ & $J_{1}^{2}=-I$\tabularnewline
\hline 
$J_{2}^{T}AJ_{2}=\bar{A}$ & (a) & (b) & (c)\tabularnewline
\hline 
$i\left(J_{2}A-\bar{A}J_{2}\right)\leq0$ & (d) & (e) & (f)\tabularnewline
\hline 
\end{tabular}
\end{table}

First, when the inequality in Eq.~(\ref{eq:ContractionCondition})
becomes equality, we will have two antilinear symmetries: $J_{1}^{T}AJ_{1}=J_{2}^{T}AJ_{2}=\bar{A}$.
Under this circumstance our result goes back to the known results
based on symmetry considerations\citep{wei_majorana_2016,li_majorana-time-reversal_2016}.
In this case parameters actually live in the maximal subgroup of the
semigroup. Many models in practical studies fall into this case, which
can be simulated by quantum Monte Carlo without the sign problem\citep{hirsch_discrete_1983,hirsch_two-dimensional_1985,lang_monte_1993,wu_exact_2003,capponi_current_2004,wu_sufficient_2005,meng_quantum_2010,hohenadler_correlation_2011,zheng_particle-hole_2011,hohenadler_quantum_2012,berg_sign-problemfree_2012,tang_berezinskii-kosterlitz-thoules_2014,huffman_solution_2014,li_solving_2015,wang_split_2015,li_majorana-time-reversal_2016,schattner_competing_2016}.
Below we list a few important examples.

(a)The negative-$U$ Hubbard models, the positive-$U$ Hubbard models
at half-filling on bipartite lattices\citep{meng_quantum_2010}, and
the Kane-Mele-Hubbard model\citep{hohenadler_correlation_2011,zheng_particle-hole_2011,hohenadler_quantum_2012}
at half-filling can all be regarded as good examples of this case.
For those models, $J_{1}$ could either be symmetric or skew-symmetric,
due to the high symmetry of the systems.

(b)A class of interacting spinless fermion models on bipartite lattices
at half-filling have been shown to be sign-problem-free using the
fermion bag approach\citep{chandrasekharan_fermion_2010} for the
continuous-time quantum Monte Carlo (CTQMC) method\citep{huffman_solution_2014}.
They have also been treated without the sign problem using the AFQMC
method under the Majorana fermion basis\citep{li_solving_2015}, and
using the CTQMC method under the framework of the split orthogonal
group\citep{wang_split_2015}. Actually our result applies to several
different kinds of QMC methods, such as CTQMC, despite their differences
in practice. That class of spinless fermion models is made up of typical
examples of the case with symmetric $J_{1}$. Another example is a
model for helical topological superconductors with interactions\citep{li_majorana-time-reversal_2016}.
Actually this case has been included in Theorem 2 of Ref.~\citep{wei_majorana_2016},
where the matrices $A$, $J_{2}$, and $J_{1}J_{2}$ play the roles
of $V$, $S$, and $P$ there, respectively.

(c)For the case with skew-symmetric $J_{1}$, the related fermion
lattice models have Kramers time-reversal invariance \citep{wei_majorana_2016,li_majorana-time-reversal_2016}.
Applications can also be found in high-spin interacting fermion systems,
e.g., the nuclear shell model\citep{lang_monte_1993} and the high-spin
Hubbard model\citep{wu_exact_2003,wu_sufficient_2005}. This sign-problem-free
property of Kramers time-reversal-invariant models also has applications
in the research of high-temperature superconductors\citep{capponi_current_2004,berg_sign-problemfree_2012,schattner_competing_2016,xu_competing_2021}.
This case has also been included in Theorem 2 of Ref.~\citep{wei_majorana_2016},
where the matrices $A$, $J_{2}$, and $iJ_{1}J_{2}$ play the roles
of $V$, $S$, and $P$ there, respectively.

Second, when the inequality in Eq.~(\ref{eq:ContractionCondition})
is not an equality, parameters actually live in the whole semigroup.
Models of this case can also be simulated by quantum Monte Carlo without
the sign problem.

(d)We note that for some models, both symmetric and skew-symmetric
$J_{1}$ are suitable. Those models correspond to the intersection
of the two semigroups. A generalized Kane-Mele-Hubbard model with
staggered magnetic field, considered in Ref.~\citep{wei_majorana_2016},
can be seen as an example.

(e)When $J_{1}$ is symmetric, the parameter region given by Eqs.~(\ref{eq:RealStructure})
and~(\ref{eq:ContractionCondition}) coincides with the result obtained
from Majorana reflection positivity. To see this clearly, one may
choose a Majorana fermion basis such that $J_{1}=\sigma_{1}\otimes I_{N}$
and $J_{2}=i\sigma_{2}\otimes I_{N}$. The fermion degrees of freedom
are grouped into two parts under this new basis $\gamma=\left(\begin{array}{c}
\gamma^{\left(1\right)}\\
\gamma^{\left(2\right)}
\end{array}\right)$, while the condition of reflection symmetry is given by Eq.~(\ref{eq:RealStructure}),
and the condition of positivity is ensured by Eq.~(\ref{eq:ContractionCondition}).
We have $\gamma^{T}A\gamma=\gamma^{\left(1\right)T}B\gamma^{\left(1\right)}+\gamma^{\left(2\right)T}\bar{B}\gamma^{\left(2\right)}+2i\gamma^{\left(1\right)T}C\gamma^{\left(2\right)}$,
with block matrices $B$ and $C$. $C$ is positive semidefinite Hermitian
matrix. This can be immediately compared to the related definitions
in Ref.~\citep{wei_majorana_2016}. All the models studied by the
fermion bag approach and the split orthogonal group approach can also
be treated by Majorana reflection positivity. 

The set of operators with Majorana reflection positivity is closed
under operator multiplication\citep{jaffe_characterization_2016,wei_majorana_2016},
which accounts for the semigroup property. Each strict contraction
element corresponds to a strictly positive operator in the sense of
Majorana reflection positivity. We mention that this strict reflection
positivity can also be used to show the uniqueness of the ground state
for finite systems\citep{lieb_two_1989,wei_ground_2015}.

(f)When $J_{1}$ is skew-symmetric, the result in the previous section
implies new sign-problem-free models. For convenience in practical
applications, we reexpress our result for this $J_{1}$ skew-symmetric
case in terms of the complex fermion basis. Without losing generality,
we can choose the Majorana fermion basis $\gamma=\left(\begin{array}{c}
\gamma^{\left(1\right)}\\
\gamma^{\left(2\right)}\\
\gamma^{\left(3\right)}\\
\gamma^{\left(4\right)}
\end{array}\right)$ so that the two skew-symmetric orthogonal matrices have the form
$J_{1}=\sigma_{x}\otimes i\sigma_{y}\otimes I_{N/2}$ and $J_{2}=-i\sigma_{y}\otimes I_{2}\otimes I_{N/2}$.
Then we define the complex fermion basis as $c_{l}=\left(\gamma_{l}^{\left(1\right)}+i\gamma_{l}^{\left(2\right)}\right)/2$
and $d_{l}=\left(\gamma_{l}^{\left(4\right)}+i\gamma_{l}^{\left(3\right)}\right)/2$,
where $l=1,\dots,N/2$ labels different components. There is a one-to-one
correspondence between the coefficient matrices $A$ which satisfy
the conditions in Eq.~(\ref{eq:RealStructure}) and the fermionic
quadratic forms with Kramers time-reversal invariance,

\begin{eqnarray}
h & = & \frac{1}{4}\gamma^{T}A\gamma=h^{\left(0\right)}+h^{\left(p\right)},\label{eq:ComplexFermionBasis}\\
h^{\left(0\right)} & = & \left(c^{\dagger},\ d^{\dagger}\right)M\left(\begin{array}{c}
c\\
d
\end{array}\right)-\left(c,\ d\right)M^{T}\left(\begin{array}{c}
c^{\dagger}\\
d^{\dagger}
\end{array}\right),\label{eq:ParticleConservingPart}\\
h^{\left(p\right)} & = & \left(c,\ d\right)RK\left(\begin{array}{c}
c\\
d
\end{array}\right)-\left(c^{\dagger},\ d^{\dagger}\right)SK\left(\begin{array}{c}
c^{\dagger}\\
d^{\dagger}
\end{array}\right).\label{eq:ParingPart}
\end{eqnarray}
Here, $M$, $R$, and $S$ are complex coefficient matrices, and $K=i\sigma_{y}\otimes I_{N/2}$.
$RK$ and $SK$ are skew-symmetric, in accordance with the fermion
anticommutation relations. The time reversal operation is given by
the unitary transformation $K$ followed by a complex conjugation
of the coefficients under the complex fermion basis. It is not difficult
to check that $K^{T}MK=\bar{M}$, $K^{T}RK=\bar{R}$, and $K^{T}SK=\bar{S}$.
Then the condition in Eq.~(\ref{eq:ContractionCondition}) is now
converted to two inequalities for the Hermitian matrices $R$ and
$S$, $R\geq0$ and $S\geq0$ under this complex fermion basis. The
particle number conserving part $h^{\left(0\right)}$ corresponds
to the generator of the maximal subgroup of the contraction semigroup,
while the pairing term $h^{\left(p\right)}$ corresponds to the invariant
cone.

Consider a Kramers time-reversal-invariant effective band Hamiltonian
defined on an arbitrary lattice, with time-reversal symmetry that
satisfies $K^{2}=-I$. We add an attractive on-site Hubbard-$U$ term
to the Hamiltonian. With appropriate HS transformations to decouple
the interaction term\footnote{To treat some interaction terms like the Hubbard-$U$ term, it might
be useful to notice that $(1-2n)=\exp\left(i\pi n\right)$ for any
fermion number operator $n$ with eigenvalues $0$ and $1$. That
kind of interaction term can be expressed by the exponential of quadratic
forms. }, sign-problem-free AFQMC simulations can be carried out for this
type of model\citep{wu_sufficient_2005}. Now we can extend this model
by adding a new pairing term that satisfies the sign-problem-free
conditions to study the proximity effect of superconductivity to topological
matters with correlation effects. Actually, by particle-hole transformation
one can also map an attractive interaction term to a repulsive one,
or a pairing term to a hopping term to study more physical problems
in strongly correlated electron systems. Those possibilities have
not been shown by any previous research.

As an example of this case, consider the model Hamiltonian $H=H_{0}+H_{\perp}+H_{p}+H_{U}$
defined on a square lattice, where
\begin{equation}
H_{0}=-t\sum_{\left\langle i,j\right\rangle }\left(c_{i}^{+}c_{j}+h.c.\right)-t\sum_{\left\langle i,j\right\rangle }\left(d_{i}^{+}d_{j}+h.c.\right),
\end{equation}
\begin{equation}
H_{\perp}=-t_{\perp}\sum_{i}\left(-1\right)^{x_{i}+y_{i}}\left(c_{i}^{+}d_{i}+h.c.\right),
\end{equation}
\begin{align}
H_{p} & =\sum_{i}[\Delta(c_{i}^{+}d_{i+\delta_{x}}^{+}-c_{i}^{+}d_{i+\delta_{y}}^{+}\nonumber \\
 & +c_{i}^{+}d_{i-\delta_{x}}^{+}-c_{i}^{+}d_{i-\delta_{y}}^{+})+h.c.],
\end{align}
\begin{equation}
H_{U}=U\sum_{i}\left(c_{i}^{+}c_{i}-\frac{1}{2}\right)\left(d_{i}^{+}d_{i}-\frac{1}{2}\right).
\end{equation}
Here $t$, $t_{\perp}$, and $U$ are real parameters, $U\geq0$,
and $\Delta$ is a complex parameter. Here $H_{\perp}$ describes
the effect of a staggered magnetic field along the $x$ axis, and
$H_{p}$ describes the $d$-wave BCS pairing. The whole Hamiltonian
depicts the proximity effect between superconductivity and antiferromagnetism
and can be simulated by the AFQMC method without the sign problem.
This can be proved by introducing the particle-hole transformation
$d_{i}\rightarrow\left(-1\right)^{x_{i}+y_{i}}d_{i}^{+}$ and suitable
HS transformations\citep{hirsch_discrete_1983}. When $t_{\perp}=0$,
the single-body term of the Hamiltonian is gapless and the whole Hamiltonian
has been used to study the quantum criticality of the chiral Heisenberg
universality class\citep{otsuka_dirac_2020}. When $t_{\perp}\neq0$,
the whole Hamiltonian exhibits a single-particle gap\citep{grossman_robust_2023}.

For many models the system contains several different kinds of degrees
of freedom. Suppose each subsystem satisfies sign-problem-free condition,
e.g., with its own choice of matrices $J_{1}$ and $J_{2}$. If the
coupling terms between the two sign-problem-free subsystems are carefully
selected, the whole system can still be sign-problem-free. In that
case, our sign-problem-free conditions are to be applied to each building
block of the whole system. This observation can be useful in the study
of multilayer systems.

For the sign-problem-free models studied in this work, the partition
function can be seen as a summation of contraction semigroup elements.
This structure can have interesting consequences, including the sign
structure of expectation values of observables. For example, for any
positive integer $m$, we have 
\begin{equation}
\textrm{tr}\left(h'_{1}h'_{2}\dots h'_{m}g_{0}\right)\geq0,\label{eq:TraceInequality}
\end{equation}
where the coefficient matrices of fermionic quadratic operators $h'_{s}$
belong to the invariant cone, $s=1,\dots,m$, and $g_{0}$ can take
any elements of the semigroup. The proof is straightforward for the
case with symmetric $J_{1}$ owing to Majorana reflection positivity.
A proof for the case with skew-symmetric $J_{1}$ using Wick's theorem
and the Kramers degeneracy theorem can be found in the Supplemental
Material. This set of inequalities also provides an alternative route
to the sign-problem-free property. They are also very useful when
considering the sign problem in zero temperature AFQMC simulations. 

\section{Conclusion And Discussion}

In this work we have presented sufficient conditions for sign-problem-free
QMC simulations of fermion lattice models. A framework based on the
concept of the semigroup has been proposed to understand this problem
in a systematic way. Sufficient conditions have been obtained, as
stated in Eqs.~(\ref{eq:RealStructure}) and~(\ref{eq:ContractionCondition}).
All previous results based on symmetry considerations and Majorana
reflection positivity can be understood well and unified naturally
within our approach. Alternative sign-problem-free models have been
constructed to show the power of our method. Such sign-problem-free
interacting fermion models share some general physical properties,
as we have demonstrated.

Although we have focused on applications in quantum lattice models
in condensed matter physics, our framework is not limited to those
cases and can also help with the sign problems in the other branches
of physics\citep{hayata_classification_2017}.

We would like to mention that the techniques used in this work can
be extended to systems with bosonic degrees of freedom.
\begin{acknowledgments}
The author would like to thank Alexander Alldridge, Lei Wang and Martin
Zirnbauer for illuminating discussions, with special thanks to Alexander
Alldridge for introducing knowledge of Lie semigroups, and Martin
Zirnbauer for the help during the revision process. We acknowledge
the financial support of the Deutsche Forschungsgemeinschaft (DFG)
under Projektnummer 316511131, TRR 183, Project A03.
\end{acknowledgments}

\bibliographystyle{apsrev4-1}
\bibliography{SGQMC}
\newpage{}

\appendix

\section*{Supplemental Material}

\section{Proof of the skew-symmetry property of $J_{2}$}

If $J_{2}$ were symmetric, for any element $A$ in the invariant
cone with metric $J_{2}$, let $A^{\prime}=\left(A+\bar{A}\right)/2$.
From $J_{2}^{T}AJ_{2}=-\bar{A}$ we know that $\left\{ J_{2},\ A^{\prime}\right\} =0$.
Then we would have
\[
\mathrm{tr}\left(J_{2}A\right)=\mathrm{tr}\left(J_{2}A'\right)=-\mathrm{tr}\left(A'J_{2}\right)=0.
\]
That is to say the only possible element in the invariant cone would
be zero matrix, which does not suit our purpose. So $J_{2}$ can only
be skew-symmetric.

\section{Proof of the anticommutation relation between $J_{1}$ and $J_{2}$}

Let $A$ be any strict contraction element in the invariant cone with
metric $iJ_{2}$, and $A^{\prime}=i\left(A-\bar{A}\right)/2$. In
this case $\left[J_{2},\ A^{\prime}\right]=0$. We also have $\left\{ J_{1},\ A^{\prime}\right\} $$=0$,
because $J_{1}^{T}AJ_{1}=\bar{A}$. We can always select an $A$ such
that the real symmetric matrix $Q=-J_{2}A^{\prime}$ is positive definite.

Now consider the real symmetric(when $J_{1}^{2}=I_{2N}$) or skew-symmetric(when
$J_{1}^{2}=-I_{2N}$) matrix
\[
X=J_{1}A^{\prime}=J_{1}J_{2}Q=-QJ_{2}J_{1}.
\]
$X$ has real orthogonal matrix $J_{1}J_{2}$ and positive real sym-
metric matrix $Q$ as its unique polar decomposition, which implies
that $\left[J_{1}J_{2},\ Q\right]=0$. Hence $\left\{ J_{1},\ J_{2}\right\} $$=0$.

\section{Proof of a series of trace inequalites}

Now we prove the inequality in Eq.~(\ref{eq:TraceInequality}) of
the main text for the case with skew-symmetric $J_{1}$. To this purpose,
we show an equivalent result that any trace with the form

\[
W=\left(-1\right)^{r}\mathrm{tr}\left[f_{4r}f_{4r-1}\dots f_{2s}f_{2s-1}\dots f_{2}f_{1}\exp\left(h\right)\right]
\]
is non-negative for any positive integer $r$, where $h$ can take
any particle number conserving fermionic quadratic forms with Kramers
time-reversal invariance. $f_{2s-1}$ ($s=1,\dots,2r$) can take any
fermion annihilation or creation operators(which we do not require
to form an orthogonal basis), while $f_{2s}$ are their images under
the time-reversal operation respectively. 

Let $Z=\mathrm{tr}\left[\exp\left(h\right)\right]\geq0$ and assume
$Z\neq0$ so that the Green's function $G$ can be defined by
\[
G_{ij}=-\frac{1}{Z}\mathrm{tr}\left\{ \left[\theta\left(i-j\right)f_{i}f_{j}-\theta\left(j-i\right)f_{j}f_{i}\right]\exp\left(h\right)\right\} 
\]
with $i,\ j=1,\dots,4r$, and $\theta$ is the unit step function.
Only the contractions between annihilation and creation operators
lead to non-zero elements of $G$. Define the skew-symmetric matrix
$\tilde{G}$ by $\tilde{G}_{ij}=G_{ij}$ when $i>j$. In the non-trivial
case there are $2r$ creation operators and $2r$ annihilation operators
in total. Therefore one can select an appropriate permuation matrix
$P$ such that
\[
\tilde{G}=P^{T}\left(\begin{array}{cc}
0 & M\\
-M^{T} & 0
\end{array}\right)P
\]
with $\det P=1$, meanwhile the Kramers time-reversal symmetry leads
to the relation 
\[
J_{M}^{T}MJ_{M}=\bar{M},
\]
where $J_{M}$ is a skew-symmetric real orthogonal matrix. So all
the eigenvalues of $M$ occur in complex conjugate pairs even when
they are real numbers, which is ensured by the Kramers degeneracy
theorem.

By Wick's theorem we have the Pfaffian expression for the expectation
\[
W=\text{\ensuremath{\frac{\left(-1\right)^{r}}{\left(2r\right)!4^{r}}}}\mathrm{pf}\left(\tilde{G}\right)Z=\frac{1}{\left(2r\right)!4^{r}}\det\left(M\right)Z,
\]
which is clearly non-negative.
\end{document}